\documentclass[conference]{IEEEtran}
\IEEEoverridecommandlockouts
\usepackage{cite}
\usepackage{amsmath,amssymb,amsfonts}
\usepackage{algorithmic}
\usepackage{graphicx}
\usepackage{textcomp}
\usepackage{xcolor}
\usepackage{mathtools}
\newtheorem{exmp}{Example}[section]
\def\BibTeX{{\rm B\kern-.05em{\sc i\kern-.025em b}\kern-.08em
    T\kern-.1667em\lower.7ex\hbox{E}\kern-.125emX}}
\begin{document}

\title{Record Linkage to Match Customer Names: A Probabilistic Approach
}

\author{\IEEEauthorblockN{Bahare Fatemi, Seyed Mehran Kazemi, and David Poole}
\IEEEauthorblockA{\textit{Department of Computer Science} \\
\textit{The University of British Columbia}\\
Vancouver, Canada \\
\{bfatemi, smkazemi, poole\}@cs.ubc.ca}
}

\maketitle

\begin{abstract}
Consider the following problem: given a database of records indexed by names (e.g., name of companies, restaurants, businesses, or universities) and a new name, determine whether the new name is in the database, and if so, which record it refers to. This problem is an instance of \emph{record linkage} problem and is a challenging problem because people do not consistently use the official name, but use abbreviations, synonyms, different order of terms, different spelling of terms, short form of terms, and the name can contain typos or spacing issues.
We provide a probabilistic model using \emph{relational logistic regression} to find the probability of each record in the database being the desired record for a given query and find the best record(s) with respect to the probabilities. 
Building on term-matching and translational approaches for search, our model addresses many of the aforementioned challenges and provides good results when existing baselines fail.
Using the probabilities outputted by the model, we can automate the search process for a portion of queries whose desired documents get a probability higher than a trust threshold.
We evaluate our model on a large real-world dataset from a telecommunications company and compare it to several state-of-the-art baselines. The obtained results show that our model is a promising probabilistic model for record linkage for names. We also test if the knowledge learned by our model on one domain can be effectively transferred to a new domain. For this purpose, we test our model on an unseen test set from the business names of the secondString dataset. Promising results show that our model can be effectively applied to unseen datasets. Finally, we study the sensitivity of our model to the statistics of datasets.
\end{abstract}


\section{Introduction}
Many companies offer services that require searching their database for a text query specified by a user. A website containing reviews for restaurants lets a user find their desired restaurant by searching its name. A website containing scientific papers lets a user find their desired paper through searching its title. A telephone company needs to search through their customer records (individual names or company titles) for customer inquiries. 

The challenge in designing a model for the purposes exemplified above arises when people abbreviate all or part of the name while the database contains the full name (e.g., searching for \emph{ICDM} when the database contains \emph{international conference on data mining}) or vice versa, change the order of the terms in the name (e.g., searching for \emph{relational probabilistic models} when the database contains \emph{probabilistic relational models}), enter only some (not all) terms in the name (e.g., searching for \emph{graphical models} when the database contains \emph{probabilistic graphical models}), shorten a long term or person's name (e.g., searching for \emph{Bayes net} when the database contains \emph{Bayesian networks} or searching for \emph{Mike Brown} when the database contains \emph{Michael Brown}), add or remove spaces (e.g., searching for \emph{drop out} when the database contains \emph{dropout}), use different spellings of terms (e.g., searching for \emph{neighbor} when the database contains \emph{neighbour}), use a common misspelling of a term (e.g., searching for \emph{busyness} when the database contains \emph{business}), and have typos.

Record linkage \cite{fellegi1969theory}, \cite{christen2012data} refers to the problem of recognizing records in two separate files which represent identical persons, or objects. It has been previously studied independently by researchers in several areas under various names including object identification \cite{tejada2002learning}, entity resolution \cite{benjelloun2009swoosh}, identity uncertainty \cite{pasula2003identity}, approximate matching \cite{gravano2001approximate}, duplicate detection \cite{monge1997efficient}, merge/purge \cite{hernandez1995merge}, or hardening soft information \cite{cohen2000hardening}. Applications of record linkage include citation matching \cite{giles1998citeseer}, person identification in different Census datasets \cite{winkler2006overview}, and identifying different offers of the same product from multiple retailers for comparison shopping \cite{bilenko2005adaptive}. The problem we study in this paper, finding the corresponding name(s) in a database for a text query, is an instance of record linkage when records are names. Hereafter, we refer to this problem as \emph{record linkage for names}. 

In this paper, we study the problem of record linkage for names when labeled data in the form of a set of $\langle$query name, desired name$\rangle$ pairs is available. We develop a probabilistic model for this task as a probabilistic approach facilitates the decision making process, e.g., for specifying an error tolerance and automating a portion of the queries. 
We experimented with several existing approaches and also developed a relational logistic regression (RLR) \cite{Kazemi:2014} model for this task which outperforms the existing approaches. We used RLR as it simplifies specifying and describing our model, may be extended when the dataset contains more fields, and empirically compares well to other related models \cite{kazemi2017comparing}. The components used in our model can be computed offline with a linear pass over the dataset. The time complexity of answering queries (online phase) for the proposed method is sub-linear in the number of names in the database. Performing a search with our proposed model only takes few seconds, and our model is to be used as a front-line service for the telecommunications company.

We tested our model empirically by conducting experiments on a large real-world dataset from a telecommunications company and compared our model with several state-of-the-art models. The obtained results show how our model outperforms the state-of-the-art. We also show how the probabilities outputted by our model facilitate decision making for \emph{query automation}. 

The knowledge learned through our model (a list of correlated terms) can be transferred to other domains. Transferring this knowledge is especially valuable for domains where labeled data does not exist, or for domains that the amount of labeled data (or the number of businesses in the database) is not enough for machine learning purposes. To test the effectiveness of knowledge transfer for our model, we train a model on a dataset created on Yelp businesses and university names and test it on the secondString dataset \cite{cohen2000hardening}. The obtained results show that the knowledge our model learns on one domain can be effectively transferred to new domains.

For different domains, the aforementioned challenges for record linkage for names (e.g., abbreviation, order change, misspellings, etc.) may occur at different frequencies. For instance, in a domain containing university names abbreviations may occur frequently, while on a domain containing paper titles abbreviations may be quite infrequent. We performed a sensitivity analysis on the dataset we created on Yelp businesses and university names to measure the effect of such statistical changes and analyze how our model is expected to perform and how it compares to existing approaches on new datasets with different statistics.


\section{Related Work}
Record linkage for names is a similar problem to short-text search \cite{kenter2015short,severyn2015learning}, where users search through documents containing short texts (e.g., considering only document titles). In short-text search, however, a query is to be matched to a document with the same \emph{meaning}. E.g., for a query containing the term \emph{passion}, the engine may score two documents one having the term \emph{passion} and the other having \emph{love} (almost) equally. Such scoring is, however, not sensible for record linkage for names. Nevertheless, many techniques developed for short-text search can be used for record linkage for names.

A classic approach for a text search problem is \emph{exact term matching}: matching exact terms in the query with those in records (or documents), weighting each term according to its importance. 
Well-known exact term matching  algorithms are TFIDF \cite{salton1988term} and Okapi BM25 \cite{robertson2009probabilistic}. 
An advantage of these approaches is that they are invariant to the change of order in the terms.
However, due to not being able to handle abbreviations, short forms of names, typos, spacing issues, etc., these approaches fail on a portion of queries.

Another class of approaches that can be applied to this problem are approximate string matching approaches \cite{hall1980approximate}. Well-known distance functions include Levenstein \cite{levenstein1965binary}, Monge-Elkan \cite{monge1997efficient}, and Jaro \cite{jaro1995probabilistic}.
Each of these approaches uses a distance function that measures the dissimilarity between two strings. 
Using distance functions in record linkage for names may not work in many examples when the appearances of query name and the desired name are different (e.g. searching for \emph{ICDM} when the database contains \emph{international conference on data mining}). Furthermore, these approaches usually perform poorly when the order of terms in the query and desired document differ.

Latent semantic models \cite{deerwester1990indexing,hofmann1999probabilistic,blei2003latent,dumais1997automatic,platt2010translingual} aim at improving the exact-term matching approaches by converting the query and the document to a smaller space called the semantic space, and then finding the similarities in the semantic space. 
A query and a record can have a high similarity score in the semantic space even if they do not have any terms in common. When labeled data is not available, these models use unsupervised methods, such as SVD, to carry out the conversion, in which case, the conversion is only loosely coupled to the evaluation metric for the retrieval task.
When labeled data is available, the conversion can be done supervised using, e.g., a deep neural network \cite{huang2013learning,palangi2016deep}. 
The labeled training set may be constructed by having an expert manually selecting the appropriate record, or may come from a user clicking a record after searching for a query. 

When supervised data is available, translational approaches \cite{gao2010clickthrough,gao2011clickthrough} learn a term to term translation between query and documents terms. The translations are learned using a labeled dataset. Studies show when large amounts of labeled data are available, translational models can be effective \cite{gao2010clickthrough,gao2011clickthrough}. 

The model we develop in this paper can be considered as a term matching algorithm, extended with ideas from translational models to address several issues that arise in record linkage for names including abbreviations, short forms of the names, common typos, and spacing issues.


\section{Background and Notations}
To be consistent with other scientific papers in this field (e.g., \cite{huang2013learning} and \cite{mitra2017neural}), we use the following terminologies: A \textbf{term} is a sequence of letters and digits. A \textbf{document} is a sequence of terms.
A \textbf{corpus}  is a set of documents. A \textbf{query} is also a document that we are interested in finding its duplicate. A \textbf{positive labeled set} is a set of $\langle query, document \rangle$ pairs where the document is the duplicate of the query in the corpus. A \textbf{negative labeled set} is a set of $\langle query, document \rangle$ pairs where the document is not the duplicate of the query. 

In record linkage for names, a \textbf{document} corresponds to a record, except that a record may have several fields but a document contains one field which is its text. A \textbf{corpus} corresponds to a database of records in record linkage. For two documents $D_1$ and $D_2$, $D_1 \cap D_2$ is the set of terms that are both in $D_1$ and $D_2$ and $D_1 - D_2$ is the set of terms that are in $D_1$ but not in $D_2$.


\subsection{TFIDF}
TFIDF \cite{salton1988term} is one of the most popular exact term matching algorithms. Beel et al. \cite{beel2016paper} reported that 83\% of text-based recommender systems in the domain of digital libraries use TFIDF.

Given a query $Q$, the TFIDF score for each document $D$ being the desired document is:
\begin{equation}
TF\-IDF(D, Q) = \Sigma_{T \in Q \cap D}{TF(T, D) * IDF(T)}
\end{equation}
\textbf{\textit{TF(T, D)}} stands for term frequency and is computed by counting the number of times $T$ appears in $D$.
\textbf{\textit{IDF(T)}} stands for the inverse of document frequency and measures how much information the term provides, that is, whether the term is common or rare across all documents in the corpus.
IDF aims at scaling down the importance of common terms and scaling up the importance of rare terms. There are many variants, but typically, the IDF score of a term $T$ for a corpus is computed as $log\frac{n}{DF(T)}$, where $n$ is the number of all documents in the corpus and $DF(T)$ is the number of documents that have the term $T$. 
Robertson et al. \cite{robertson2004understanding} justified this $IDF$ score information theoretically. 



For record linkage for names, the TF part of the TFIDF is usually $1$ as each term (almost always) appears at most once in a document (e.g., we rarely see the name of a company, person, or paper having one term twice). Thus, we ignore the TF part and only use the IDF.


\subsection{Relational logistic regression}
Relational logistic regression (RLR) \cite{Kazemi:2014} is the relational counterpart of logistic regression and the directed counterpart of Markov logic \cite{MLN}. We start with some terminology:

A \textbf{population} refers to a set of \textbf{objects}. The \textbf{size} of a population is a non-negative number indicating its cardinality.
\textbf{Logical variables (logvars)} start with lower-case letters, and \textbf{constants} denoting objects start with upper-case letters. Associated with a logvar \textit{x} is a population $\Delta_{x}$. A lower-case and an upper-case letter written in bold refer to a set of logvars and a set of objects respectively. 

A \textbf{parametrized random variable (PRV)} is of the form $F(t_1, ..., t_k)$ where $F$ is a k-ary function symbol and each $t_i$ is a logvar or a constant. 

A \textbf{literal} is a PRV or its negation. A \textbf{formula} is made up of literals connected with conjunction or disjunction. 
A \textbf{weighted formula (WF)} is a tuple $\langle F, w\rangle$ where $F$ is a formula and $w$ is a weight.    

We write a substitution as $\theta = 
{\langle x_1, ..., x_k \rangle / \langle t_1, ..., t_k \rangle}$ where each $x_i$ is a different logvar and each $t_i$ is a logvar or a constant in $\Delta_{x_i}$. A \textbf{grounding} of a PRV can be obtained by a substitution $\theta = {\langle x_1, ..., x_k \rangle / \langle X_1, ..., X_k \rangle}$ mapping each of its logvars $x_i$ to an object $X_i \in \Delta_{x_i}$. Applying a substitution $\theta = {\langle x_1, ..., x_k \rangle / \langle t_1, ..., t_k \rangle}$ on a formula $F$ (written as $F\theta$) replaces each $x_i$ in $F$ with $t_i$.

Let $H(\mathbf{x})$ be a PRV whose probability depends on a set $\phi$ of PRVs not including $H$. We call $\phi$ the \emph{parents} of $H$. \textbf{Relational logistic regression (RLR)} defines a conditional probability distribution for $H(\mathbf{x})$ given an assignment of truth values $\Pi$ to every ground PRV of $\phi$, using a set $\psi$ of WFs only containing PRVs from $\phi$:
\begin{equation}
P_\psi(H(\mathbf{X})=True|\Pi) = \sigma(\sum_{\langle F, w\rangle \in \psi} w * \eta(F\theta, \Pi))
\end{equation}
where $\eta(F\theta, \Pi)$ is the number of instances of $F\theta$ that are true w.r.t. $\Pi$, and $\sigma(z)=\frac{1}{1+exp(-z)}$ is the \emph{Sigmoid} function.

\begin{figure}
\includegraphics[width=\columnwidth]{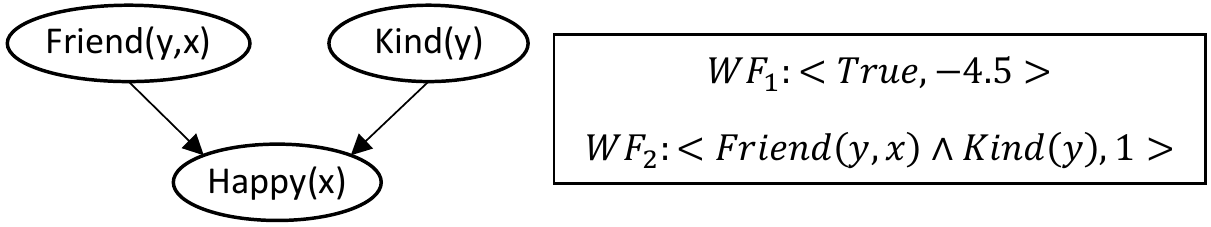}
\caption{An RLR model taken from Kazemi et al. \cite{Kazemi:2014}.}
\label{KindFriend}
\end{figure}

\begin{exmp}
Consider we want to find the probability of a person being happy and we know that happiness has a relation with the number of kind friends the person has such that the more kind friends the person has the happier he/she is. The model in Fig. \ref{KindFriend} shows our theory. In this model let $\Pi$ be an assignment of values to $Friend$ and $Kind$. 
RLR sums over $\psi=\{WF_1,WF_2\}$ resulting in:
\begin{equation}
\begin{split}
P_\psi(Happy(X)=True \mid \Pi) = \sigma(-4.5 \\ + 1 * \eta(Friend(y, X) \wedge Kind(y), \Pi))
\end{split}
\end{equation}
where $\eta(Friend(y, X) \wedge Kind(y), \Pi)$ represents the number of objects in $pop(y)$ for which $Friend(y,X) \wedge Kind(y)$ is true according to $\Pi$, corresponding to the number of friends of $X$ that are kind. When this count is greater than or equal to 5, the probability of $X$ being happy is closer to one than zero; otherwise, the probability is closer to zero than one. Therefore, the two WFs model \texttt{"}someone is happy if they have at least 5 friends that are kind\texttt{"}. Note that -4.5 and 1 are weights of two formula that are learned. 
\end{exmp}

\begin{figure}
\begin{center}
\includegraphics[width=0.6\columnwidth]{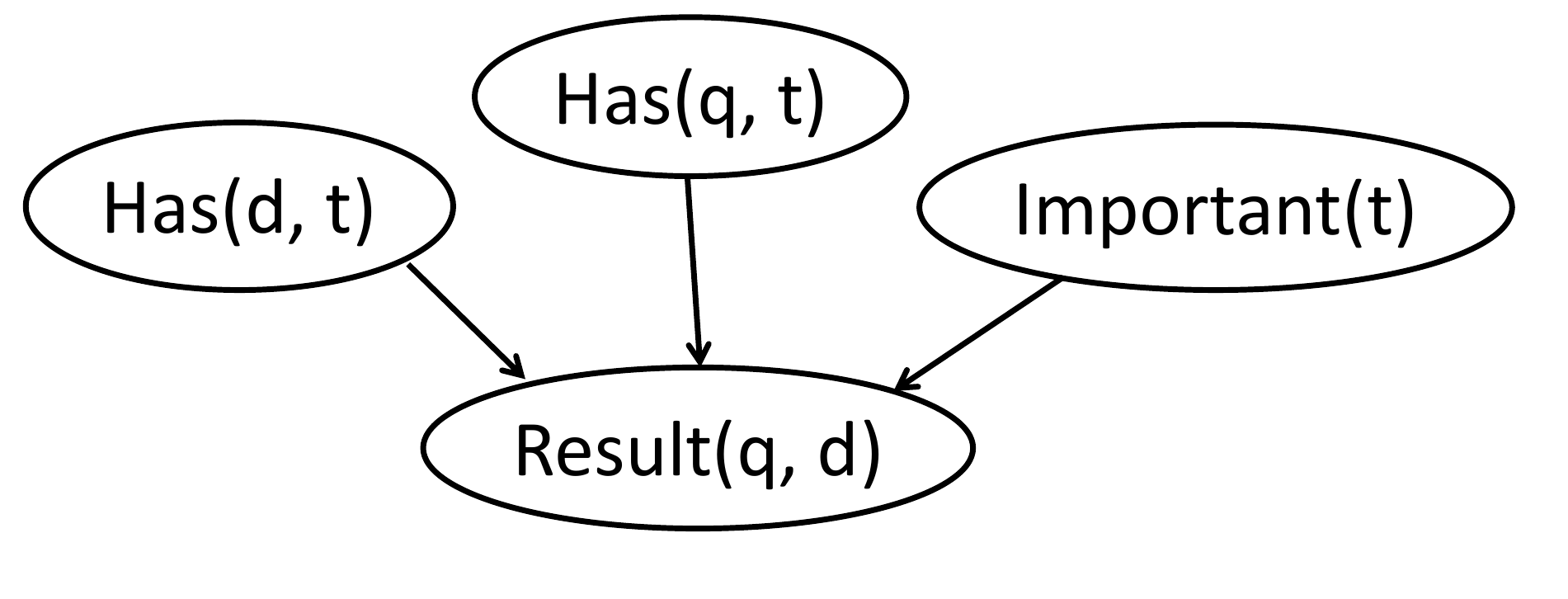}
\end{center}
\caption{An RLR model for Result(q, d)}
\label{Result}
\end{figure}

\begin{exmp}
Suppose we want to have an RLR model as in Fig.~\ref{Result} to predict $P(Result(q,d))$: the probability of document $d$ being the result of searching for a query $q$. Also, suppose we have a list of important terms. An RLR model may use the WFs $\psi=\{\langle True, -4 \rangle$, $\langle Has(d, t) \land Has(q, t) \land Important(t), 1.5 \rangle\}$
where $Has(d,t)$ is true if term $t$ is in document $d$, and $Important(t)$ is true if $t$ is important.

Let $\Pi$ be an assignment of truth values to all ground PRVs in $Has$ and $Important$.
RLR sums over the WFs in $\psi$ resulting in $P_\psi(Result(Q, D)=True \mid \Pi) =  \sigma(-4.0 + 1.5 * \eta(Has(D, t) \land Has(Q, t) \land Important(t),\Pi))$
where $\eta(Has(D, t) \land Has(Q, t) \land Important(t),\Pi)$ represents the number of objects in $\Delta_{t}$ for which
$Has(D,t) \land Has(Q, t) \land Important(t)$ is true according to $\Pi$, corresponding to the number of important terms that are both in $Q$ and $D$. 
When this count is greater than or equal to 3, the probability of $D$ being the result of $Q$ is closer to one than zero.
Therefore, \texttt{"}a document is a result of a query if they share at least 3 important terms\texttt{"}. 

\end{exmp}

Following \cite{Kazemi:2014,fatemi2016learning}, we assume w.l.o.g. that formulae in WFs have no disjunction, indicate $true$ and $false$ with $1$ and $0$ respectively, replace conjunction with multiplication, and allow atoms with continuous functions in WFs (e.g., $\langle Has(d, t) \land Has(q, t) \land Important(t), 1.5 \rangle$ is replaced with $\langle Has(d, t) * Has(q, t) * Important(t), 1.5 \rangle$).

\section{A Model of Record Linkage}

Let $t$, $q$, and $d$ be logvars corresponding respectively to terms, queries, and documents. $Has(d, t)$ is a Boolean PRV indicating whether document $d$ has $t$ or not (which is observed for all documents and terms), $IDF(t)$ is an observed real valued PRV indicating the IDF score for terms, and $Result(q,d)$ is $True$ when $d$ is the desired document for $q$.

\subsection{A probabilistic TFIDF-based model}\label{algorithm_TFIDF}
We design an RLR model to assign probabilities to each document being the desired document for a query. We start with a basic RLR model and improve it step by step. The basic RLR model defines the conditional probability of $Result(d,q)$ using the following WFs: $\langle True, w_0 \rangle$, $\langle Has(q, t) * Has (d, t) * IDF(t), w_1 \rangle$. When both instances of $Has$ are true, it contributes the weight $IDF(t) * w_1$. RLR sums over the above WFs resulting in:
\begin{equation} \label{rlr-result-eq}
P(Result(Q,D)\mid\Pi) = \sigma(w_0 + w_1 * \sum_{T\in Q \cap D}IDF(T))
\end{equation}
where $Q$ and $D$ are respectively a specific query and document.
Having a positive and a negative labeled set, the weights $w_0$ and $w_1$ can be learned using gradient descent. 




\subsection{Normalizing the basic model}\label{normalized_tfidf}

One issue with the basic model is that for a query $Q$ containing only a few terms, or only common terms, $\sum_{T\in Q \cap D}IDF(T)$ may be generally very small. For such queries, unless $w_1$ is unrealistically h igh, $w_0 + w_1 * \sum_{T\in Q \cap D}IDF(T)$ would be small even for the desired document, causing the output probability of the model to be low for such queries. Furthermore, the score of each document $D$ for a query $Q$ in the basic model depends on the \emph{number} of query terms that are in $D$, not the \emph{proportion} of them.

To address this issue we update the WFs to normalize the sum of the IDF scores by dividing it by the maximum IDF score a document can get for a given query. A document gets the maximum score for a query $Q$ if it has all the terms in $Q$. In such a case, the sum of the IDF scores is $\sum_{T\in Q} IDF(T)$. So we update the basic model to have the following WFs:
\begin{equation}\label{first_WF_tfidf}
\langle True, w_0 \rangle
\end{equation}
\begin{equation}\label{second_WF_tfidf}
\langle Has(q, t) * Has (d, t) * IDF(t) * InvSumIDF(q), w_1 \rangle
\end{equation}
Where $InvSumIDF(Q)=\frac{1}{\sum_{T\in Q} IDF(T)}$. Hereafter, we refer to the RLR model with the WFs (\ref{first_WF_tfidf}) and (\ref{second_WF_tfidf}) as the \emph{TFIDF} model.

\begin{exmp}
Consider a query $Q=[T_1, T_2, T_3, T_4]$, the document that can get the maximum score for $Q$ has all terms $T_1$, $T_2$, $T_3$, and $T_4$. $SumIDF(Q)$ in this query is:
\begin{center}
$IDF(T_1) + IDF(T_2) + IDF(T_3) + IDF(T_4)$
\end{center}
and $InvSumIDF(Q)$ is as:

\begin{center}
$\frac{1}{IDF(T_1) + IDF(T_2) + IDF(T_3) + IDF(T_4)}$
\end{center}
\end{exmp}

\begin{exmp}\label{tf-idf-example}
Consider a query $Q=[T_1, T_2, T_3, T_4]$ and a document $D=[T_1, T_2,T_5,T_6]$. Then:
\begin{equation}\label{tf-idf-example-eq}
\begin{split}
P(&Result(Q,D))=\sigma(w_0+w_1*\\
&\frac{IDF(T_1)+IDF(T_2)}{IDF(T_1)+IDF(T_2)+IDF(T_3)+IDF(T_4)})
\end{split}
\end{equation}
In which $IDF(T_1)+IDF(T_2)$ is the sum of the $IDF$ score that $D$ gets for query $Q$ and $IDF(T_1)+IDF(T_2)+IDF(T_3)+IDF(T_4)$ is the maximum $IDF$ score a document can get for query $Q$ and normalizes the $IDF(T_1)+IDF(T_2)$.
\end{exmp}

\subsection{Adding translations}
As mentioned, a TFIDF-based method may fail on documents that use different terms than those in the query. 
Let the PRV $Tr(t, t')$, where $t$ and $t'$ are two logvars with population of terms, represent the proposition that term $T\in \Delta_{t}$ is a translation (or a part of a translation) of term $T'\in \Delta_{t'}$. We will explain how such a PRV can be learned using positive and negative labeled sets in later sections. We assume translation relation is symmetric (if $t$ is a translation or a part of a translation of $t'$, then $t'$ is also a translation or a part of a translation of $t$). The following WF can use this PRV:
\begin{equation}\label{third_WF_tfidf}
\begin{split}
\langle &  Has(q, t) * Has (d,t') * \neg Has(q,t') * \neg Has(d,t) \\ & * Tr(t, t') * IDF(t) * InvSumIDF(q), w_1 \rangle
\end{split}
\end{equation}

 

This WF considers pairs of terms $\langle T, T'\rangle$ such that $T \in Q$, $T \notin D$, $T' \in D$, $T' \notin Q$, and $T'$ is a part of translation for $T$ with probability $Tr(T,T')$, and gives $Tr(T, T') * IDF(T)$ score to the document. This WF complements the WF in~(\ref{second_WF_tfidf}). If a term $T$ in $Q$ is also in $D$, then the WF in~(\ref{second_WF_tfidf}) gives $IDF(T)$ score to $D$. If $T$ is not in $D$  but there exists a term $T'$ in $D$ (which is not in $Q$) that can be a translation of $T$ with probability $Tr(T,T')$, then the new WF gives $Tr(T,T') * IDF(T)$ score to $D$. That is because even though $D$ does not have the exact term $T$, with probability $Tr(T,T')$ it has a term that corresponds to $T$. Since this WF is complementing the WF in~(\ref{second_WF_tfidf}), we used the same weight $w_1$.
\begin{exmp}
In Example~\ref{tf-idf-example}, with the same query $Q$ and document $D$ suppose $Tr(T_3, T_5)=0.9$. Then the numerator of the fraction in Eq.~\ref{tf-idf-example-eq} will be summed with $0.9 * IDF(T_3)$. Note that if $Tr(T_1, T_6)=0.8$, we do not add $0.8 * IDF(T_1)$ to the score as the document contains $T_1$ as well and we have already given $IDF(T_1)$ score to the document. 
\end{exmp}

\subsection{1-to-many and many-to-1 translations}
For some $T_i$ that is in $Q$ but not in $D$ and $T_j$ that is in $D$ but not in $Q$ such that $Tr(T_i, T_j)>0$, $T_i$'s translation may contain multiple terms and $T_j$ may be only one term in the translation of $T_i$. As an example, if $Q$ is \emph{ICDM}, $D$ is \emph{international conference on DM} and $Tr(\emph{ICDM}, \emph{international})>0$, \emph{international} is only one term of the translation for \emph{ICDM}; the other terms are \emph{conference}, \emph{on}, and \emph{DM}.

The IDF formulation makes the strong assumption that each term appears in a document independently of the other terms \cite{robertson2004understanding}. Therefore, if $Q$ is \emph{ICDM} and $D$ is \emph{International conference on DM}, our current WFs will give scores to $D$ for all its terms independently. This is, however, not intuitive as the terms in $D$ in such cases are highly dependent: a document containing the terms \emph{international}, \emph{conference} and  \emph{on} is much more likely to have the term \emph{DM} than a random document. 

To address this issue, when we learn the values of the $Tr$ PRV, we also compute for each term $T$ the $MaxTr(T)$ as $max_{D\in C}\sum_{T' \in D} 1_{Tr(T,T')>0}$ where $1_{Tr(T,T')>0}$ is $1$ if $Tr(T,T')>0$ and $0$ otherwise and $C$ is the set of all documents in the corpus $c$. This number corresponds to the maximum number of terms $T'$ in a document for which $Tr(T,T')>0$. Assuming $InvMaxTr(t)$ is a PRV whose value for each $T \in t$ is $\frac{1}{MaxTr(T)}$, we update the WF in \ref{third_WF_tfidf} as:
\begin{equation}\label{fourth_WF_tfidf}
\begin{split}
\langle &  Has(q, t) * Has (d,t') * \neg Has(q,t') * \neg Has(d,t) * Tr(t, t') \\ & * InvMaxTr(t) * IDF(t) * InvSumIDF(q), w_1 \rangle
\end{split}
\end{equation}

We refer to the RLR model with WFs (\ref{first_WF_tfidf}), (\ref{second_WF_tfidf}), and (\ref{fourth_WF_tfidf}) as \emph{TFIDF+TR}.
\begin{exmp}
We expect that $MaxTr($\emph{ICDM}$) = 5$ (for \emph{international}, \emph{conference}, \emph{on}, \emph{data}, and \emph{mining}) and
$MaxTr($\emph{center}$) = 1$ (for, e.g., \emph{centre}).

\end{exmp}

\begin{exmp}
In Example~\ref{tf-idf-example} with the same query and document, suppose we have:
$Tr(T_3, T_5)=0.9$, $Tr(T_3, T_6)=0.8$, and $InvMaxTr(T_3)= \frac{1}{3}$. 
Then the numerator of fraction in Eq.~\ref{tf-idf-example-eq} will be summed with $\frac{0.9}{3} * IDF(T_3) + \frac{0.8}{3} * IDF(T_3)$. The denominator is same as before because it should represent the maximum score a document can get for this query for normalization and remain unchanged with this extra information about translation pairs.
\end{exmp}

\subsection{Learning $Tr(t,t')$}
We learn and populate $Tr$ for each pair $\langle T,T' \rangle$ of terms using a regularized proportion which we threshold to give a sparse representation. To define our regularized proportion, we first provide some intuition.

Suppose $\langle Q, D\rangle$ is a pair in the positive labeled set. For two terms $T$ and $T'$, if $T \in Q - D$ and $T' \in D - Q$ (or $T' \in Q - D$ and $T \in D - Q$) then $T$ might be a term in the translation of $T'$. Such occurrences are positive events regarding $Tr(T,T')$. 

Now consider the case where $T \in Q$, $T \in D$ and $T' \in D$. This case reduces the possibility of $T'$ being a term in the translation of $T$ as $T$ also appears in $D$. Therefore, such occurrences are negative events regarding $Tr(T, T')$. 

A natural way to compute $Tr(T, T')$ is by dividing the number of positive events by the sum of the number of positive and negative events.
Let: Match$(T, T')=\sum_{\langle Q,D \rangle \in PLS} Has(Q, T) * Has(D, T') * \neg Has(Q, T') * \neg Has(D, T)$ and Seen$(T, T')=\sum_{\langle Q,D \rangle \in PLS} Has(Q, T) * Has(D, T')$.
Then we let $Tr(T, T') = \frac{Match(T, T') + c_1}{Seen(T, T') + c}$, where $c_1$ and $c$ are pseudo-counts and are learned by cross-validation. 
Pseudo-counts impose a prior that a pair of terms are less likely to be part of the translation of each other, unless we see them match multiple times. The pseudo-counts allow small amounts of data to have some influence, but not much, whereas large amounts of data can overwhelm the prior.

\subsection{Adding bigrams}
Spacing is an important challenge in record linkage for names: the query may contain a space between two terms where the document does not (e.g., searching for \emph{drop out} where the document contains \emph{dropout}) or vice versa. In order to handle such cases, we use the bigrams of the query and the documents where a bigram is a concatenation of two consecutive terms in the query or documents.

There are two cases that need to be considered: 

\textbf{1) Query contains the bigram:}\label{Query contains the bigram} Consider a document $D=[T_1, T_2, \dots, T_n]$. If a query contained the term $concat(T_1,T_2)$, then $T_1$ and $T_2$ are parts of the translation of $concat(T_1,T_2)$. 
As an off-line process which will be done once, before the queries arrive, for each document $D=[T_1, T_2, \dots, T_n]$ we set $Tr(concat(T_{i}, T_{i + 1}), T_{i})=Tr(concat(T_{i}, T_{i + 1}), T_{i + 1})=1$ for $i: 1 \leq i < n$ if $concat(T_{i}, T_{i + 1})$ appears in at least one document in the dataset\footnote{We observed that by only considering the bigrams that appear in at least one document, the number of bigrams stored in the $Tr$ matrix decrease substantially while the accuracy is not affected much.}. This allows us to recognize the two elements of the bigram in the document that appear in the query. Note that we do not add $Tr(T_{i}, concat(T_{i}, T_{i + 1}))$ and $Tr(T_{i + 1}, concat(T_{i}, T_{i + 1}))$ to the $Tr$ matrix (which can be helpful when a document contains the bigram), as adding them to the matrix causes each term $T_i$ to have many potential translations in the $Tr$ matrix thus slowing down the search. Instead, we handle the case where document contains the bigram with a different approach as described below.
\begin{exmp}
Suppose in the corpus $C$ there is a document $D$ as \emph{drop out}. As explained, we set $Tr(dropout, drop) = 1$ and $Tr(dropout, out) = 1$. So if we search for a query Q = \emph{dropout}, the translation pairs helps us to find the components of the bigram \emph{dropout}, which are \emph{drop} and \emph{out}.
\end{exmp}
\textbf{2) Document contains the bigram:} Suppose there is a document $D$ and $T_i$ and $T_j$ are two consecutive terms in the query $Q$, and neither of $T_i$ and $T_j$ appear in the document. Then we should also look for the concatenation of these two terms, $concat(T_i,T_j)$, in the document. In order for our WFs to remain unaffected, if $D$ contains neither $T_i$ nor $T_j$ but it contains $concat(T_i,T_j)$, we assume $D$ also has $T_i$ and $T_j$.

We refer to the \emph{TFIDF+TR} model after adding the bigrams \emph{TFIDF+TR+BG}. 

\begin{exmp}
Suppose we want to search for $Q$ = \emph{drop out}. The document $D$ = \emph{dropout} will get the score of a document that has both terms \emph{drop} and \emph{out} for having the bigram \emph{dropout}.
\end{exmp}



\begin{figure*}
\begin{center}
\includegraphics[width=2\columnwidth]{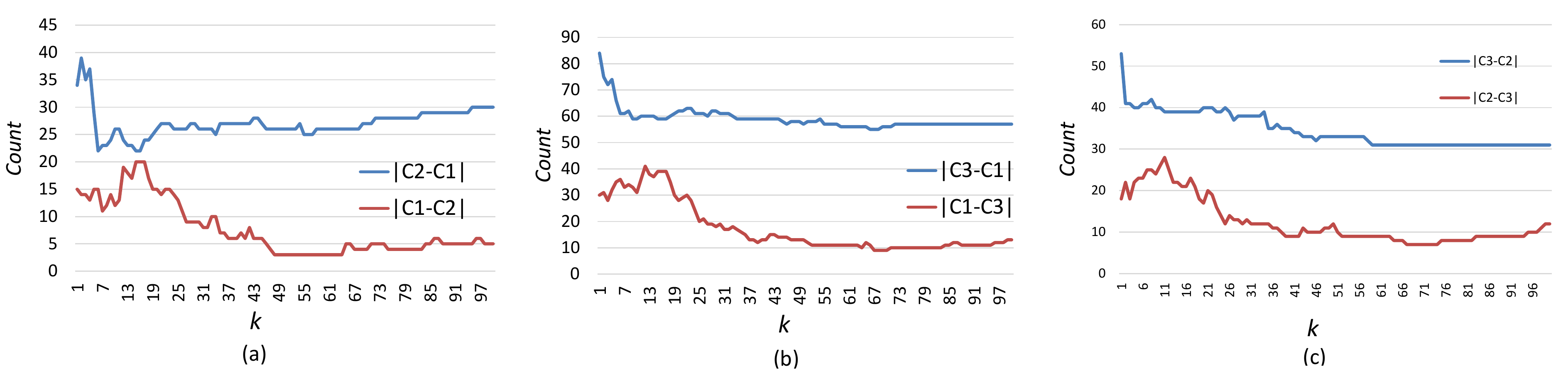}
\end{center}
\caption{Set differences for the \emph{hit@k} of \emph{TFIDF} (C1), \emph{TFIDF+TR} (C2), and \emph{TFIDF+TR+BG} (C3).}
\label{tfidf-vs-others}
\end{figure*}

\subsection{Implementation}
In order to implement our model efficiently, we use the following data structures. For the WF in~(\ref{second_WF_tfidf}), given a query $Q$, we need to score documents $D$ that have at least a term in common with $Q$. We use an \textbf{inverted index} which is a hash map from terms to the documents that contain that term. Then for each $T \in Q$, we can access the documents having $T$ and update their scores. For the translations,  the $Tr$ matrix can be very large and so needs to be implemented efficiently. We store a $TrSet$ hash-map from terms to the set of terms that may be in their translation together with the corresponding probabilities. Then for each term $T \in Q$, we retrieve all documents not containing $T$ but containing a term $T' \in TrSet(T)$ and update their score according to the probabilities. As explained, when we construct a bigram $concat(T_i, T_j)$ in our offline process, we add the key $concat(T_i, T_j)$ (with values for $T_i$ and $T_j$) to our $TrSet(T)$ only if the term $concat(T_i, T_j)$ exists in at least one document. This reduces the size of $TrSet$ and the running time substantially. These models are all developed for a front-line application and produce results in few seconds for a large dataset.

\section{Empirical Results} 


In our experiments, we aim at answering three questions. 
\begin{itemize}
\item \textbf{Q1:} How does \emph{TFIDF+TR+BG} compare to other existing approaches for record linkage for names and how helpful \emph{TR} and \emph{BG} are?
\item \textbf{Q2:} Can we effectively use the probabilities outputted by our models for query automation?
\item \textbf{Q3:} Is it possible to learn translations on a dataset and use them for a second dataset(i.e. transfer learning)?
\item \textbf{Q4:} How does \emph{TFIDF+TR+BG} and other approaches perform when the statistics of the dataset change?
\end{itemize}

We design four experiments to answer each of these questions.

\subsection{\textbf{Q1:} How well does \emph{TFIDF+TR+BG} perform?}
To answer \textbf{Q1}, we compare \emph{TFIDF+TR+BG}'s performance on a private dataset with several well-known benchmarks in the literature.

\textbf{Dataset:}
Our dataset contains a list of business names of the customers of a telecommunications company. The company offers a service which requires searching a name entered by a customer in the dataset. In this application, the list of customer names is the corpus, each stored customer name is a document, and any searched customer name is a query. 

The telecommunications company dataset contains approximately 650K customer names. Each month 4K different customers (queries) are searched. Currently, except for some obvious cases, the final document for a query is found or endorsed by an expert, providing a large positive labeled set with approximately 1600K pairs. We used the pairs in our positive labeled set corresponding to queries until a certain point in time as training data and the queries in the next two months (almost 8K queries) as test data. This makes the task an extrapolation task (predicting the future) that should be more challenging than interpolation.
We also created a negative labeled set for learning the weights of the model by pairing a query with 5 documents other than the desired document following Huang et al. \cite{huang2013learning}. Note that output probabilities of the model will change if we use a number other than 5 but the ranking will not. 

\textbf{Learning $\textbf{Tr}$:}
In order to learn the translation PRV $Tr(t, t')$, we found on a validation set that pseudo-counts $c_1=1$ and $c=5$ give good results. To sparsify this PRV and make its size manageable, we only keep the pairs of terms whose probability was at least 0.7, where 0.7 is also selected on a validation-set. This provided us with approximately 10K pairs of terms.
Table~\ref{tokens-table} represents some pairs of terms learned by our regularized proportion algorithm. The translations we learned through our regularized proportion algorithm can be transferred and used for other similar tasks.

\begin{table}[t]
\small
  \caption{Some pairs of terms learned by our regularized proportion for $Tr(t,t')$: each two terms in a columns shows a learned pair.}
  \label{tokens-table}
  \centering
  \begin{tabular}{c|c|c|c}
    
    Association & Service & Centre & Consulting              \\
    
    Assoc & Svc & Center & Consultin              \\
    Assn & Srvc & Centr & Consul              \\
    Associate & Srvs & Cntr & Consltng              \\
    Asso & Srv & Cntre & conslt               \\

  \end{tabular}
\end{table}

We also tried the heuristic proposed in \cite{gao2010clickthrough} which in our formulation can be written as 
$Tr(T, T') = \frac{Seen(T,T')}{QF(T')}$, where $QF(T')$ corresponds to the number of queries in the positive labeled set that have term $T'$. Accepting only the pairs with at least $0.7$ probability, this heuristic provided approximately 500K pairs of terms which severely slowed down the search engine. Furthermore, we found that the pairs generated using this heuristic are not suited for our task, and found the main reason to be the use of $Seen(T,T')$ in the numerator. While having $Seen(T, T')$ in the numerator may be sensible for short-text search, the following example shows why it may not be suitable for record linkage for names. 

\begin{exmp}
Let $Q=[T_1, T_2, T_3]$ and the corresponding document $D=[T_1, T_2, T_4]$ be a pair of $\langle Q, D \rangle$ in the positive labeled set. According to the heuristic proposed in \cite{gao2010clickthrough}, $\langle Q, D \rangle$ pair supports $Tr(T_1, T_2)$ and $Tr(T_1, T_4)$, i.e. $T_1$ being part of the translation for $T_2$ and $T_4$. However, since $T_1$ also appears in $D$, not only this pair should not support $Tr(T_1, T_2)$ and $Tr(T_1, T_4)$, but also it should reduce these probabilities. That is because, if $T_2$ and $T_4$ were parts of the translation of $T_1$, then $T_1$ would not have appeared in $D$.
\end{exmp}



\begin{figure}
\begin{center}
\includegraphics[width=0.9\columnwidth]{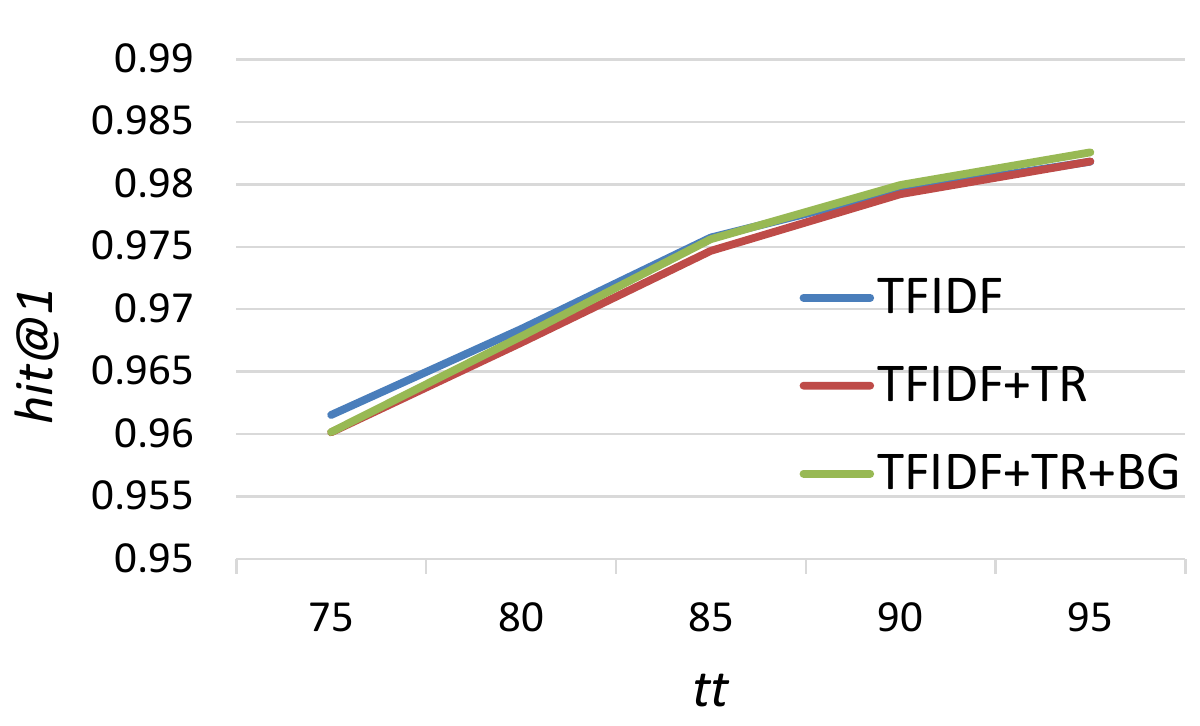}
\end{center}
\caption{\textit{hit@1} of the three methods \emph{TFIDF}, \emph{TFIDF+TR}, and \emph{TFIDF+TR+BG} for different trust thresholds \textit{tt}.}
\label{auto1}
\end{figure}

\begin{figure}
\begin{center}
\includegraphics[width=0.9\columnwidth]{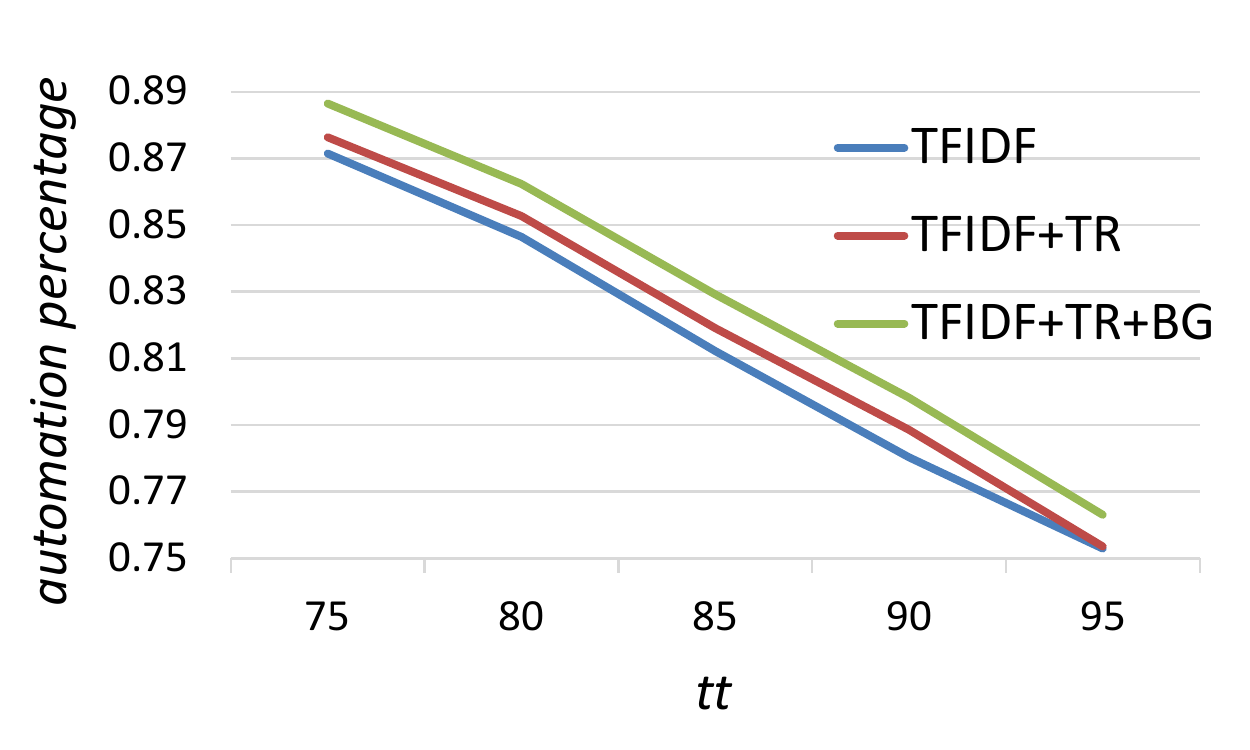}
\end{center}
\caption{\textit{automation percentage} given the trust threshold \textit{tt} for the three methods \emph{TFIDF}, \emph{TFIDF+TR}, and \emph{TFIDF+TR+BG}.}
\label{auto2}
\end{figure}

\textbf{Baselines:} 
We compare our models to several baselines. \emph{Exact match} refers to matching the query to a document with the exact name. \emph{Shared-terms} scores documents based on the number of terms they have in common with the query. \emph{Levenstein} and \emph{Jaro-Winkler} are two well-known distance functions and are widely used for approximate string matching. For speeding these distance functions up, we only score documents that have at least one term in common with the query. 
\cite{huang2013learning} propose several deep learning architectures for search. \emph{L-WH DNN} is the architecture that outperforms the other deep architectures as well as other latent semantic models in \cite{huang2013learning}'s experiments.

We learn the weights of all these model using the positive and negative labeled sets. We break the ties by picking the document that has fewer terms.

For each query in our test set, we score the documents using these methods and pick the top $k$. Following \cite{bordes2013translating,nguyen2017overview}, for a specific value of $k$, we define \emph{hit@k} as the percentage of test queries whose desired document appears in the top $k$ retrieved documents.


\textbf{Results:} Table~\ref{baseline-table} shows the \emph{hit@k} for some values of \emph{k} for our models and several baseline methods on the telecommunications company dataset.

The results in Table~\ref{baseline-table} show that 57.83\% of the test set are exact matches. Distance functions do not work well in our application since in many cases, the string of the correct match is very different than the string of the searched query (e.g., as in the \emph{ICDM} vs \emph{international conference on DM}). Another reason why methods based on distance functions do not work well is that they do not consider the importance of terms. As an example consider searching for the query \emph{ICDM association}, where the database contains two documents 1- \emph{ICDM}, and 2- \emph{NIPS association}. In this case, the second document has a lower edit distance than the first document, while a document having \emph{ICDM} is a better document than one having \emph{association} for this query. Ignoring the importance of the terms misleads algorithms towards selecting the second document. 

According to the results, deep neural network models also do not work well in our application. We found the reason to be that many terms in the customer and company names are unique and only appear a few times, therefore there is not enough data for these model to learn appropriate weights for these terms. Note that for record linkage for names, a term appearing fewer times is usually more important and carries more weight. The state-of-the-art deep learning models for information retrieval rely on learning embeddings for terms \cite{mitra2017neural}. With only a few occurrences of many terms, learning appropriate embeddings for terms is difficult.

\begin{table}[t]
\small
  \caption{\emph{hit@k} for 4 values of $k$ on the telecommunications company dataset. The winner is in bold.}
  \label{baseline-table}
  \centering
  \begin{tabular}{c|c|c|c|c}  
   &\emph{hit@1}&\emph{hit@5}&\emph{hit@10}&\emph{hit@100}\\ \hline
    exact-match & 57.83 & 57.83 & 57.83 & 57.83\\
    shared-terms & 88.76 & 92.32 & 93.19 & 95.53\\
    Levenstein & 84.16 & 87.23 & 88.16 & 91.29\\
    Jaro-Winkler & 88.35 & 91.76 & 92.68 & 95.31\\
    L-WH DNN & 75.90 & 80.11 & 81.93 & 88.74\\
    \emph{TFIDF} & 91.33 & 95.24 & 95.95 & 97.32\\
    \emph{TFIDF+TR} & 91.58 & 95.42 & 96.13 & 97.64\\
    \emph{TFIDF+TR+BG} & \textbf{92.03} & \textbf{95.63} & \textbf{96.31} & \textbf{97.88}\\
  \end{tabular}
\end{table}

Table \ref{baseline-table} also demonstrates that both translations and bigrams have a positive effect on the performance of our model in terms of \emph{hit@k}.

In order to better demonstrate the effect of translations and bigrams, for some value of $k$, let $C_1=C_{TFIDF}$, $C_2=C_{TFIDF+TR}$, and $C_3=C_{TFIDF+TR+BG}$ be the set of test queries for which the desired document is in the top $k$ retrieved documents
when using \emph{TFIDF}, \emph{TFIDF+TR}, and \emph{TFIDF+TR+BG} respectively. As $k$ grows, these sets either do not change or grow in size. However, the difference between them may differ for different values of $k$.

Fig.~\ref{tfidf-vs-others} represents the set differences between $C_1$, $C_2$, and $C_3$. It can be viewed that for all values of $k$, $|C_2-C_1|$ is always bigger than $|C_1-C_2|$ and similarly for $C_3$ and $C_1$, indicating that adding translations and bigrams always helps improve the \emph{hit@k}. Note that as $k$ becomes larger, $|C_2-C_1|$ and $|C_3-C_1|$ become larger and the set differences $|C_1-C_2|$ and $|C_1-C_3|$ become quite small (5-10 queries out of the 8000 test queries). This shows that our proposed methods cover almost everything \emph{TFIDF} covers. That is because, for queries with different terms than the actual document, \emph{TFIDF} cannot find the desired document even for a very large $k$, whereas the other two methods may be able to do so.

The results in Table~\ref{baseline-table} and Fig.~\ref{tfidf-vs-others} answer to our \textbf{Q1}. They indicate that \emph{TFIDF+TR+BG} performs well empirically outperforming state-of-the-art approaches. They also show that translations and bigrams both have a positive effect on the accuracy of the model. 


\subsection{\textbf{Q2:} Automating searching for a query}
Given that our model outputs probabilities for documents being the desired document for a query, we can associate it with a utility function. One possible utility function to be combined with this model is to set a trust threshold $tt$ such that if the top output of the model has a probability more than $tt$, then it is considered as the desired document without having an expert examine it, i.e. automating the process for a percentage of the queries. There are two important criteria for picking $tt$: \begin{itemize}
\item \textbf{\emph{hit@1:}} if we trust the top outputs of a model that passes the threshold, what percentage of them will be the actual desired documents.
\item \textbf{\emph{automation percentage:}} if we trust the top outputs of a model that passes \emph{tt}, what percentage of queries will be automatically matched to a document, without expert verification.
\end{itemize}

Fig.~\ref{auto1} represents the \emph{hit@1} of our three models for different values of \emph{tt}. It can be seen that the charts for the three models overlap and they all have similar performances and high \emph{hit@1} when \emph{tt} is set high enough. Fig.~\ref{auto2} shows the automation percentage vs. \emph{tt}. It shows both translations and bigrams increase the automation percentage for every value of \emph{tt} that we tried.

Fig.~\ref{auto1} and Fig.\ref{auto2} provide the answer to \textbf{Q2}. They show that the probabilities outputted by our models can be effectively used for query automation. They also show that both translations and bigrams improve the automation percentage thus providing more evidence for \textbf{Q1}.

\subsection{\textbf{Q3:} Transfer learning}
While the telecommunications company dataset is large and contains many positive pairs, in many similar applications (e.g., for smaller companies) such a dataset may not be available. In such cases, it is possible to learn the translations and bigrams over other datasets and use them for the dataset at hand (i.e. transfer learning). That is because the translations and bigrams are, for the most part, domain-independent.

In order to empirically test the transferability of translations and bigrams and answer \textbf{Q3}, we conduct an experiment in which we find the translations and bigrams on one dataset and then use \emph{TFIDF+TR+BG} in a new domain containing unseen business names.

\textbf{Dataset:} We collected a set of 130K business names from Yelp and 500 university names and their abbreviations from Wikipedia. We also collected a set of equivalent names, terms with more than one spellings and common misspellings from web\footnote{This set is only used for generating a labeled set. We will not use these in testing.}. Using the collected datasets, we generated a positive labeled dataset to train our model on. In order to generate train pairs, we use the following procedure. For each name in our dataset, we generate $K$ positive pairs (i.e. $K$ duplicates) where $K$ comes from a normal distribution with mean $\mu$ and standard deviation $\sigma$. Each duplicate is generated as follows. With probability $p_{change}$, some change is being applied to the duplicate and with probability $1-p_{change}$ the duplicate is the same as the original name. If a change is to be made to the duplicate name, with probability $p_{abbreviation}$ the whole name is abbreviated. If the abbreviation is not applied to the name, with probability $p_{some}$ one term in the duplicate name is removed from the name. The term to be removed is selected randomly with probability proportional to the frequency of the term (more frequent terms are more likely to be removed). With probability $p_{equivalence}$, one of the terms in the duplicate name is replaced with one equivalent form of it (e.g., \emph{center} may be replaced with \emph{centre}). With probability $p_{space}$, one of the spaces in the duplicate name is selected randomly and removed from the name. With probability $p_{order}$, two terms are selected from the duplicate name randomly and are swapped. Finally, with probability $p_{typo}$, a random typo is introduced in one of the terms in the duplicate name. We select these probabilities to be similar to the telecommunications company dataset to make them more reflect the real-world. We consider our validation set as a dataset of 9K pairs generated with this process. The test set is a completely unseen dataset from a different source than the train data source. The test set is a set of 600 business names from secondString dataset \cite{cohen2000hardening}. 


\textbf{Results:} We learn translations and bigrams over the collected dataset similarly to the previous section. Then we test all baselines on the secondString dataset. The results are available in Table~\ref{baseline-table-2}.
\begin{figure*}
\begin{center}
\includegraphics[width=2\columnwidth]{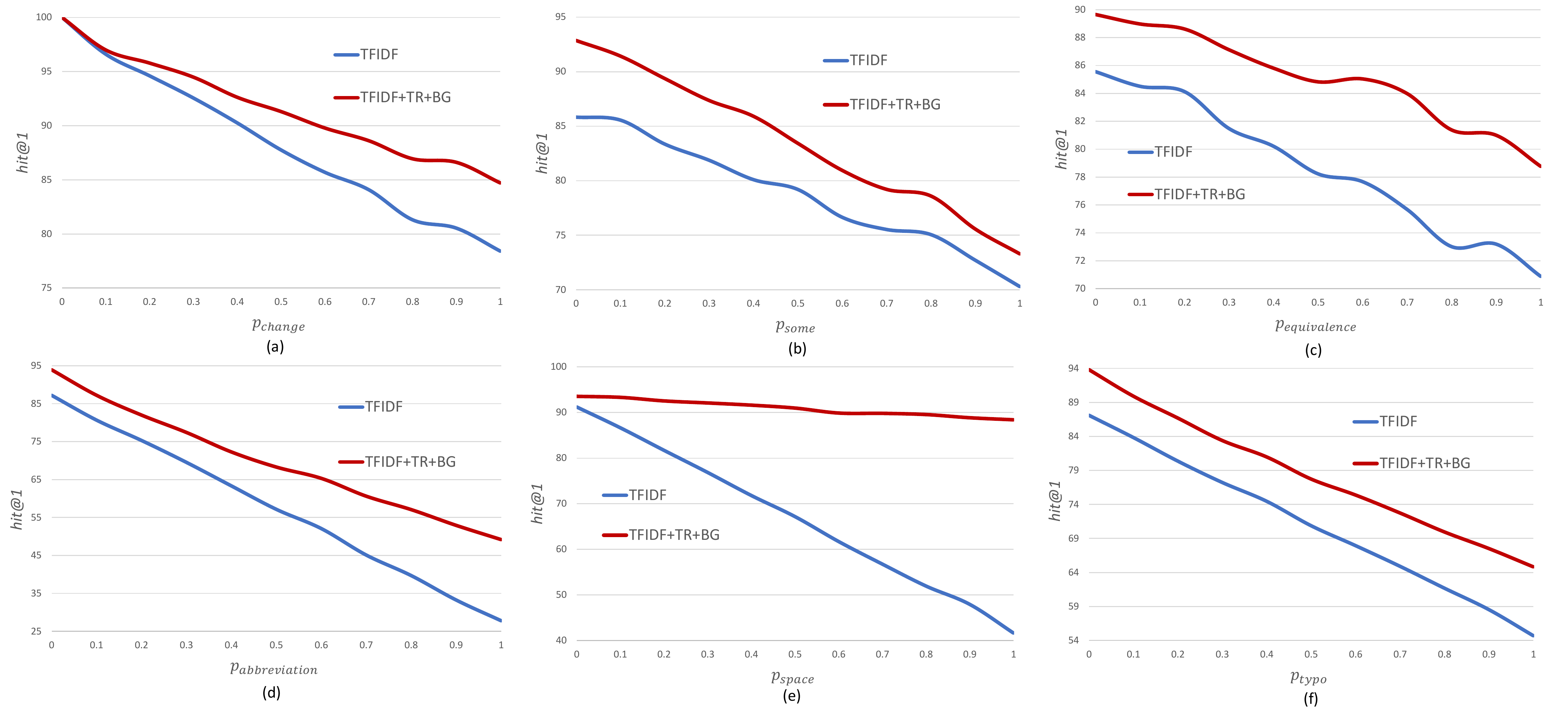}
\end{center}
\caption{\textit{hit@1} for \emph{TFIDF} and \emph{TFIDF+TR+BG} when (a) $p_{change}$, (b) $p_{some}$, (c) $p_{equivalence}$, (d) $p_{abbreviation}$, (e) $p_{space}$, and (f) $p_{typo}$ varies.}
\label{all}
\end{figure*}
\begin{table}[t]
\small
  \caption{\emph{hit@k} for 4 values of $k$ on the business names of secondString dataset. The winner is in bold.}
  \label{baseline-table-2}
  \centering
  \begin{tabular}{c|c|c|c|c}  
   &\emph{hit@1}&\emph{hit@5}&\emph{hit@10}&\emph{hit@100}\\ \hline
    exact-match & 23.63 & 23.63 & 23.63 & 23.63\\
    shared-terms & 81.98 & 90.08 & 93.38 & 96.36\\
    Levenstein & 80.66 & 88.92 & 93.71 & 96.03\\
    Jaro-Winkler & 87.76 & 93.05 & 94.21 & 95.86\\
    L-WH DNN & 38.01 & 44.62 & 47.60 & 68.59\\
    \emph{TFIDF} & 88.42 & 93.38 & 94.04 & 96.36\\
    \emph{TFIDF+TR} & 94.21 & 94.71 & 95.04 & 97.19\\
    \emph{TFIDF+TR+BG} & \textbf{95.54} & \textbf{94.87} & \textbf{95.20} & \textbf{97.35}\\
  \end{tabular}
\end{table}

According to the results, one can see that the translations and bigrams learned over a dataset can be helpful for new datasets as \emph{TFIDF+TR+BG} performs better than \emph{TFIDF}. This experiment answers \textbf{Q3} affirmatively positively. It also provides more evidence for answering \textbf{Q1} as \emph{TFIDF+TR+BG} performs better than the other baselines.

 

\subsection{\textbf{Q4:} Sensitivity to dataset statistics}
In order to answer \textbf{Q4}, we conduct experiments to show the sensitivity of \emph{TFIDF} and \emph{TFIDF+TR+BG} to dataset statistics. For this purpose, we generated datasets from the Yelp businesses and university names as described earlier, each time varying one of the parameters. We randomly split the generated data into training and testing sets, train our models on the training set and testing its performance on the testing set. Each of the following charts shows \textit{hit@1} of \emph{TFIDF} and \emph{TFIDF+TR+BG} when all the other parameters are fixed except one:
\begin{itemize}
\item Fig.~\ref{all}(a) shows the \textit{hit@1} of \emph{TFIDF} and \emph{TFIDF+TR+BG} when $p_{change}$ varies. As $p_{change}$ increases, the dataset becomes more and more challenging and the \textit{hit@1} of both methods decrease. However, it can be viewed that \emph{TFIDF} is affected more severely than \emph{TFIDF+TR+BG} and the gap between the two methods gradually increases. This shows that the translations and bigrams play a positive role in making our model learn about the variations of the duplicate names and help make more accurate predictions.

\item In Fig.~\ref{all}(b), the \textit{hit@1} of \emph{TFIDF} and \emph{TFIDF+TR+BG} are plotted when $p_{some}$ varies. The chart shows increasing $p_{some}$ makes \textit{hit@1} of both methods to decrease almost equally. That is because when more terms are being removed from the query name, finding the desired document becomes harder for both methods.

\item In Fig.~\ref{all}(c), Fig.~\ref{all}(d), and Fig.~\ref{all}(e), the \textit{hit@1} of \emph{TFIDF} and \emph{TFIDF+TR+BG} are plotted when $p_{equivalence}$, $p_{abbreviation}$, and $p_{space}$ vary respectively. The charts show that as $p_{equivalence}$ or $p_{abbreviation}$ or $p_{space}$ increase, the \textit{hit@1} of \emph{TFIDF} is more severely affected than \emph{TFIDF+TR+BG}. That is again because of the positive role the translations and the bigrams play in learning about the equivalences, abbreviations, and spacing issues.
\item In Fig.~\ref{all}(f), the \textit{hit@1} of \emph{TFIDF} and \emph{TFIDF+TR+BG} are plotted when $p_{typo}$ varies. The chart shows that both \emph{TFIDF} and \emph{TFIDF+TR+BG} are not able to handle random typos.
\end{itemize}
 
According to the charts, \emph{TFIDF+TR+BG} is expected to be an effective model for datasets with more variations in the duplicate names in terms of equivalent names, common misspellings, abbreviations, and spacing issues. This shows that our model is robust across several types of variations in the dataset. However, \emph{TFIDF+TR+BG} is not expected to work well on datasets where typos occur very frequently. Extending this model to better address typos is an interesting direction for future research.







\section{Conclusion}
In this paper, we studied an instance of record linkage problem for names.  
We developed a probabilistic model using relational logistic regression. We started with a probabilistic TFIDF-based model, then we added the possibility of recognizing two terms that are not identical but may be part of the translation of each other and also addressed the spacing issues. We tested our models on a large dataset from a telecommunications company and compared with several baselines. Obtained results indicated that our model outperforms existing state-of-the-art baselines. We showed that the knowledge learned in our model can be transferred to new domains. We also analyzed the sensitivity of our model to variations in the dataset and showed that our model is robust across several variations.
\bibliographystyle{IEEEtranS}
\bibliography{bib_file}
\end{document}